\documentclass[journal]{IEEEtran}

\usepackage[dvips]{graphicx}

\graphicspath{{figures/}}
\DeclareGraphicsExtensions{.eps}

\usepackage[cmex10]{amsmath}

\usepackage{amssymb}

\usepackage{cite}

\begin{document}

\title{Probing the sheath electric field using thermophoresis in dusty plasma. \\Part I: Numerical analysis}

\author{Victor~Land, Erica~Shen, Matthew~Benesh, Lorin~Matthews, Truell~Hyde~\IEEEmembership{Fellow,~IEEE}
\thanks{V. Land, L. Matthews, and T. Hyde are with the Center for Astrophysics, Space Physics and Engineering Research, at Baylor University, Waco, TX, 76798-7316 USA, e-mail: victor\_land@baylor.edu,  (see http://www.baylor.edu/CASPER).}
\thanks{E. Shen participated in Baylor University's High School Scholar Summer Research Program}
\thanks{M. Benesh was formerly with Baylor and is currently at Cambridge.}%
\thanks{Manuscript received xxxxx xx, 2009; revised xxxxx xx, 2009.}}

\markboth{IEEE transactions on plasma science,~Vol.~XX, No.~XX, XXXXX~2010}%
{Land \MakeLowercase{\textit{et al.}}: Dust particles as probes for the sheath electric field by applying additional
thermophoresis}

\maketitle

\begin{abstract}
By combining the results from a self-consistent 2D dusty plasma fluid model and a 3D N-body code, the equilibrium position
and crystal structure were determined for dust particles levitated in
the sheath in a modified Gaseous Electronics Conference (GEC) reference cell, in which the lower electrode was heated or cooled. 
The Debye length, charge and electric field were reconstructed on a sub-millimeter scale by applying a previously
developed, independent method. However, this method seems to
overestimate the charge, and hence underestimate the electric field. Even corrected for this fact, the results show that the dust is levitated on the plasma side of the
Bohm point. The ion drag, which is not fully taken into account, probably plays an important role in the force balance.
\end{abstract}


\IEEEpeerreviewmaketitle

\section{Introduction}

\IEEEPARstart{T}{he} sheath in a confined plasma is the volume where the nearly constant potential in the bulk is connected to
the potential on the walls. Strong electric fields repel electrons from and attract ions to the walls. Hence,
the form of the sheath electric field is of interest for plasma applications, like surface coating, deposition and
etching, and other applications in the automotive, microchip, and solar cell industries \cite{Kuhr2003, vanzantbook, Shah1999}.

In dusty plasma experiments, performed for instance in a modified GEC reference cell, like
the one discussed in
this paper, these electric fields levitate dust against gravity. The vertical force balance and the radial balance between a confining potential 
and the inter-particle forces typically result in thin two-dimensional dust crystals. These crystals allow the study of many solid state phenomena, such as waves and phonons, melting and solidification, super-diffusion, Mach
cones, and turbulence on spatial and temporal scales accessible with ordinary optical techniques \cite{Nunomura2002,Morfill1999,Liu2008,Nosenko2002}.

The properties of these systems depend on the dust charge and the Debye length, which depend on
the properties of the sheath. Measurement of the electric field using probes is too disruptive to
obtain reliable results, while optical measurements close to the electrode (either by passively observing plasma emission \cite{Huebner2009}, or by using some form of
induced fluorescence) are technically difficult \cite{Mahony1997}. 

Fortunately, dust particles themselves can act as probes in the sheath, basically being tiny floating Langmuir probes. Many 
\emph{dust-as-probe} techniques have been employed in the past: oscillating particles by applying an additional
low-frequency potential, manipulating particles with lasers \cite{Zafiu2001}, or adding perturbations to the plasma (for 
instance a sudden change in DC bias) have all been examined. Other studies used the dust levitation height to
determine the sheath edge, observing multiple particles of different sizes simultaneously to
determine the sheath electric field profile \cite{Samarian2001}.

There are many difficulties with these techniques. First, measuring the vertical force balance requires knowledge of
\emph{both} the dust charge and the electric field and a combination of techniques is required to obtain the
electric field profile. Secondly, in the case of sheath-edge determinations, additional techniques such as plasma
emission observation or probe measurements are required to justify the results, since it is not clear \textit{a
priori} why the dust should float at the sheath edge. Finally, many techniques
depend on perturbing the plasma. Since the plasma and the dust are necessarily coupled (especially for many dust particles), 
this adds uncertainty to the measurements.

In this paper, we discuss a method to obtain the dust charge and Debye length for a dust crystal levitated in the sheath of
a modified GEC cell,
confined radially by a parabolic potential created by a shallow circular cutout in a plate placed on top of the lower electrode, by using top-view
images of the crystal and measuring the radius and inter-particle distances, as explained in \cite{Hebner2002}. Once the
charge is obtained, we proceed to trace the sheath electric field
using thermophoresis to adjust the equilibrium height of the dust, a method which does \emph{not} depend on
perturbing the plasma. Results obtained with a combination of a self-consistent dusty plasma fluid model and a N-body code
are employed for this discussion. 

\section{Dust charge determination}\label{sec:theory}

Following \cite{Hebner2002}, we assume that the potential above the cutout in a plate placed on top of the powered electrode has a parabolic radial
dependence, and write the potential as a function of the height with a radial shift,

\begin{eqnarray}
\phi &=& \phi(z-h(r)),\\
h(r) &=& cr^2,
\end{eqnarray}

\noindent where the constant $c$ will be defined later. 

Next, we consider the vertical force balance on dust particles with mass $m_D$ and charge $q_D$ located in the sheath. The
forces acting on the particles
include the upward electrostatic force due to the vertical sheath electric field, the force of gravity, with acceleration
$g$, the
thermophoretic force, $F_{th}$, due to the applied heating or cooling of the lower electrode (thermophoretic forces without electrode
heating or cooling are negligible), and finally the ion drag force, $F_{id}$, due to ions accelerated downwards in the
sheath:

\begin{equation}\label{eq:verticalbalance}
m_D \ddot{z} = -m_D g - q_D \frac{d\phi(z-h(r))}{dz} + F_{th} - F_{id} = 0.
\end{equation}

\noindent The radial electrostatic force is considered to be a harmonic restoring force that stabilizes the dust crystal against the
repulsive inter-particle interactions, which gives
\begin{equation}
m \ddot{r} = -q_D \frac{d\phi(z-h(r))}{dr} \equiv -k r.
\end{equation}

\noindent Using partial differentiation and defining the prime to denote the partial derivative with respect to the argument
of a function, we can use equation \ref{eq:verticalbalance} to rewrite the radial electric field. This gives

\begin{equation}
-q_Dh^{'}(r){\phi}^{'}(z-h(r))=h^{'}(r)\left[m_Dg -F_{th} + F_{id}\right].
\end{equation}

\noindent Inserting the definition for $h(r)$, we find for $k$

\begin{equation}\label{eq:keq}
k = 2c\left[m_Dg -F_{th} + F_{id}\right],
\end{equation}

\noindent which means that we have expressed the coefficient for the \emph{radial} restoring force in terms of the
\emph{vertical} equilibrium forces and $c$. 

The dust particles suspended in the sheath will form a crystal lattice. The equation of state for a crystal
consisting of $N$ particles, with inter-particle spacing $\Delta$, interacting through an interaction potential $V(\Delta)$
is given by \cite{Hebner2002}

\begin{equation}
P = -\frac{1}{N}\frac{d\left[3NV(\Delta)\right]}{d(\sqrt{3}\Delta^2/2)} - \frac{\sqrt{3}}{\Delta}\frac{dV(\Delta)}{d\Delta}.
\end{equation}

Using appropriate boundary conditions, two equations can be derived relating the radius where the inter-particle
distance goes to infinity, $R_{\infty}$, and the central inter-particle
spacing ${\Delta}_0$ to the Deybe length, ${\lambda}_D$, and the dust charge:

\begin{equation}\label{eq:Req}
R_{\infty}^2 = \frac{3}{k}\left(3+\frac{{\Delta}_0}{{\lambda}_d}\right)V({\Delta}_0),
\end{equation}

\noindent and

\begin{equation}\label{eq:Neq}
N = \frac{2 \pi \sqrt{3}}{k {\Delta}_0}\left(\frac{1}{{\Delta}_0}+\frac{1}{{\lambda}_d}\right)V({\Delta}_0).
\end{equation}

\noindent In the above, $V$ is assumed to be a screened Coulomb potential $V(r) = q_D^2
\exp(-r/{\lambda}_D)/4\pi{\epsilon}_0r$. $R_{\infty}$ is related to the crystal radius, $R_M$, and the outer
inter-particle spacing, $s_M$, through $R_{\infty} \approx R_M + s_M \sqrt{3/2}$. Solving for $V({\Delta}_0)$ and substituting in equation
\ref{eq:Req} gives an equation for the Debye length:

\begin{equation}\label{eq:lambda}
{\lambda}_D = {\Delta}_0\left[\frac{A-S}{3S-A}\right],
\end{equation}

\noindent where we have defined the total crystal surface area $A = \pi R_{\infty}^2$ and the total surface area covered by $N$
Wigner-Seitz cells measured at the center: $S = \sqrt{3}{\Delta}_0^2N/2$. Using this equation for ${\lambda}_D$, we find for the dust charge:

\begin{equation}\label{eq:charge}
q_D = \sqrt{\frac{4\pi{\epsilon}_0{\Delta}_0kR_{\infty}^2}{3\left(3+\frac{3S-A}{A-S}\right)\exp(\frac{A-3S}{A-S})}}.
\end{equation}

Measuring the central inter-particle spacing, the maximum radius of the dust crystal, the outer
inter-particle spacing, and the total number of
particles in the crystal allows us to determine the Debye length. We can then derive the dust charge provided we have
established the value for $k$.

In order to determine $k$, we need to know the vertical equilibrium forces, as well as
$c$. In our model the melamine-formaldehyde (MF) particles are monodisperse spheres. This means we know the
gravitational force. The only significant thermophoretic force is the one we apply by heating/cooling the lower
electrode. For particles with a diameter below
4 microns, the ion drag can be significant in the sheath \cite{Hebner2002}. In our simulations we use 2 micron diameter
particles. The fluid model self-consistently solves for the ion drag, although the N-body code does not include the effect of the
ion drag. For the current analysis, we therefore neglect the effect of the ion drag, although we will discuss its possible
importance later.

As mentioned, the radial confinement potential is provided by a cylindrical depression in the lower electrode. $c$ is
determined by the geometry of this cutout: the narrower and deeper
the cutout, the steeper the potential well. In \cite{Hebner2002} spherical cutouts were used, so that the value of $c$ could
be directly related to the radius of curvature of the cutout $R_c$ through $c=0.5/R_c$. Following this approach, we
approximate the cylindrical cutout by the sphere that exactly touches the lowest point of the cutout as well as the edge. The radius of the 
modeled cutout is 12.5 mm, and it is 0.5 mm deep. From $R_c \sin(\theta) =
12.5$ mm and $R_c \cos(\theta) = R_c-0.5$ mm, we find an effective radius of curvature for our cutout of 160 mm. 
We thus use $k=\left(m_Dg - F_{th}\right)/0.16$ $[N/m]$.

\section{Numerical models}\label{sec:models}

The fluid model self-consistently solves for the coupled plasma and dust parameters, but is unable to resolve the
inter-particle interactions on the microscopic level and hence provides no information on the crystal structure. The N-body
code solves the inter-particle interactions, thus providing the crystal properties, but the plasma properties are not
self-consistently solved. The fluid model is therefore used in this study to provide the external confinement, as well as the input
parameters for the N-body code, for given discharge settings. We briefly discuss both models; a complete description can be
found in the references provided in the sections below.

\subsection{N-body code: \emph{box-tree}}

\emph{Box-tree} \cite{Richardson1993} integrates the equations of motion for particles moving in a simulation \emph{box}
under
prescribed external forces, while calculating the interaction between the particles using a \emph{tree}-method employing multipole
expansions for the electrostatic forces. The box and the external forces can be set to represent for instance dust particles in Saturn's
F-Ring\cite{Matthews2004}, or particles in a laboratory crystal \cite{Qiao2003}. In this study, the external forces
considered include the electrostatic force of
the sheath, gravity and themophoresis. Since the plasma properties are not determined by \emph{box-tree}, the electric field, dust charge, and
Debye length were set using values obtained from the fluid model. Dust particles were then introduced with random
positions (but zero initial velocity), and their kinetic energy was dissipated through a drag force representing the neutral
drag until an equilibrium crystal was formed.

\subsection{The fluid model}

The fluid model employed \cite{Land2006} solves the continuity equations for the electrons and positive argon ions, as well as for the electron energy
density, assuming a drift-diffusion equation for the fluxes. The sources and sinks for the electrons (and electron energy density) include excitation of argon atoms, ionization
and recombination on the dust particles. The ions are assumed to locally dissipate their energy in collisions with the
neutral atoms. Therefore, an explicit equation for the ion energy density is not solved. The electric field is found from
Poisson's equation, including the dust charge density.  These equations are initially iterated on sub-RF timescales without the presence of dust, until the solution has become 
periodic over a RF-cycle. Then, the addition of the dust is simulated by adding source terms for the dust, below the upper electrode. 

The transport of the dust fluid is solved by assuming a balance between the neutral drag and the other forces, which allows for
a drift-diffusion type equation for the dust. Gravity is a constant force which only depends on the dust size. The electrostatic force is calculated from the time averaged electric field and
the dust charge, which is calculated from the local plasma parameters using Orbital Motion Limited (OML) electron- and ion currents, including the effect of
charge-exchange collisions on the ion current \cite{Allen1992,Ratynskaia2004}. 

The ion drag force is calculated from the local ion flux interacting with the
dust particles. The
ion collection cross-section is derived from OML theory. The ion scattering cross-section includes the effects of scattering
beyond the Debye length \cite{Khrapak2002}, the anisotropic screening caused by significant ion drift \cite{Hutchinson2006},
and the effect
of charge-exchange collisions \cite{Ivlev2005}. 

In order to compute the thermophoretic force, the neutral gas temperature profile is
calculated by iterating the power balance. The sources include the dissipation of energy by the ions, as well as heating
through atoms impinging on the hot dust particle surfaces. In order to find the dust particle surface temperature, a balance is solved between the recombination
of ions and electrons on the surface on the one hand, and the thermal radiation of the dust particles and the conduction to
the gas on the other hand. The temperature of the surrounding walls and electrodes sets the boundary conditions. These can be
changed to include heating or cooling of surfaces. 

\subsection{Approach}

The known variables are the particle mass, the effective radius of our cutout, the applied temperature gradient and hence the thermophoretic force, calculated
through 

\begin{equation}\label{eq:thermophoresis}
F_{th} = -\frac{32}{15}\frac{{\kappa}_Ta^2}{v_T}\nabla T_{gas},
\end{equation}

\noindent with ${\kappa}_T = 0.01772$ the heat conduction coefficient for argon, $a$ the particle radius, and $v_T$ the
thermal velocity of the background gas. We therefore know $k$, since we are not concerned with the ion drag force right
now. 

Given a set of discharge parameters, i.e. the pressure and input power, we obtain the Debye length, the dust charge, and
the electric field from the fluid model. These are then used as input in
the box-tree model. Once the dust crystal has reached equilibrium, the observables are obtained from the crystal (i.e. the number
of particles in the crystal, the crystal radius, the central and outer inter-particle spacing) and the method of
\cite{Hebner2002} discussed above is used to reconstruct the Debye length, charge and electric field. By varying the
thermophoretic force, the charge and electric field profiles are obtained throughout the sheath region. This provides a
method to determine whether or not this results in an acceptable reconstruction of the electric field, and also if using thermophoresis allows us to probe the
electric field throughout the sheath.

\section{Results}\label{sec:modelresults}

Here we present the results for an argon plasma at 200 mTorr pressure and 2 Watts of absorbed power in the geometry
of a modified GEC cell. We introduce 1000 two micron diameter MF particles, while varying the temperature on the lower,
powered electrode. 

\begin{figure}[!h]
\centering
\includegraphics[width=2.5in]{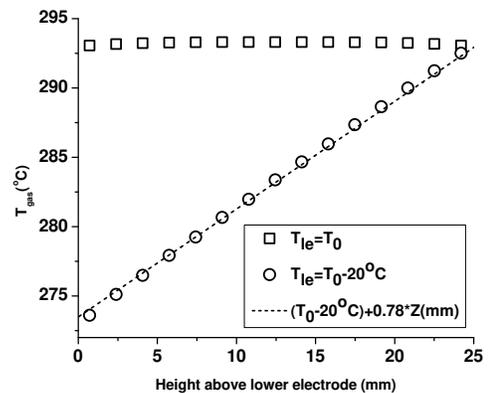}
\caption{The vertical temperature profile above the center of the electrode for the case where the lower electrode is at
room-temperature, and for the case where it is cooled to 20${}^{\circ}$ C below room-temperature.}
\label{temperatures}
\end{figure}

In the calculation of the thermophoretic force in \emph{box-tree}, it is assumed that the temperature gradient is simply given by the difference
between the temperatures of the electrodes divided by the distance between the electrodes, i.e. no large gradients are
expected. Figure \ref{temperatures} shows two gas temperature profiles along the central axis, obtained with the fluid model;
one where the lower electrode was held at room-temperature, and one where the lower electrode was cooled to 20${}^{\circ}$ C
below room-temperature. We see that in the first case, the vertical temperature gradient between the electrodes is negligible and
certainly too small to create a significant thermophoretic force. The second case clearly shows that the temperature profile varies linearly with 
height and shows no local gradients; in other word, the temperature gradient is indeed simply the
temperature difference divided by the distance between the electrodes.

\begin{figure}[!ht]
\centering
\includegraphics[width=2.5in]{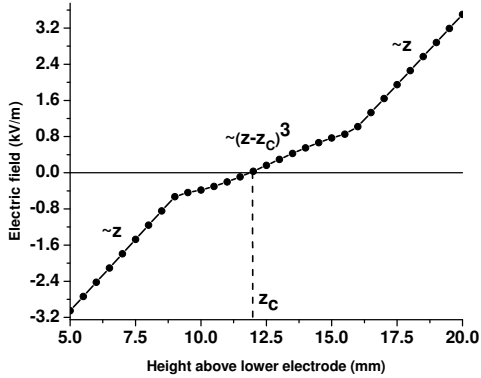}
\caption{A fit to the vertical electric field between the electrodes obtained from the fluid model for a run at 200 mTorr and
2 W of input power. The sheath electric field and the bulk electric field are clearly distinguished. The discharge does not
exhibit a real quasi-neutral bulk where the electric field completely vanishes, which has also been observed experimentally.}
\label{fit_electric_field}
\end{figure}

\begin{figure}[!ht]
\centering
\includegraphics[width=2.5in]{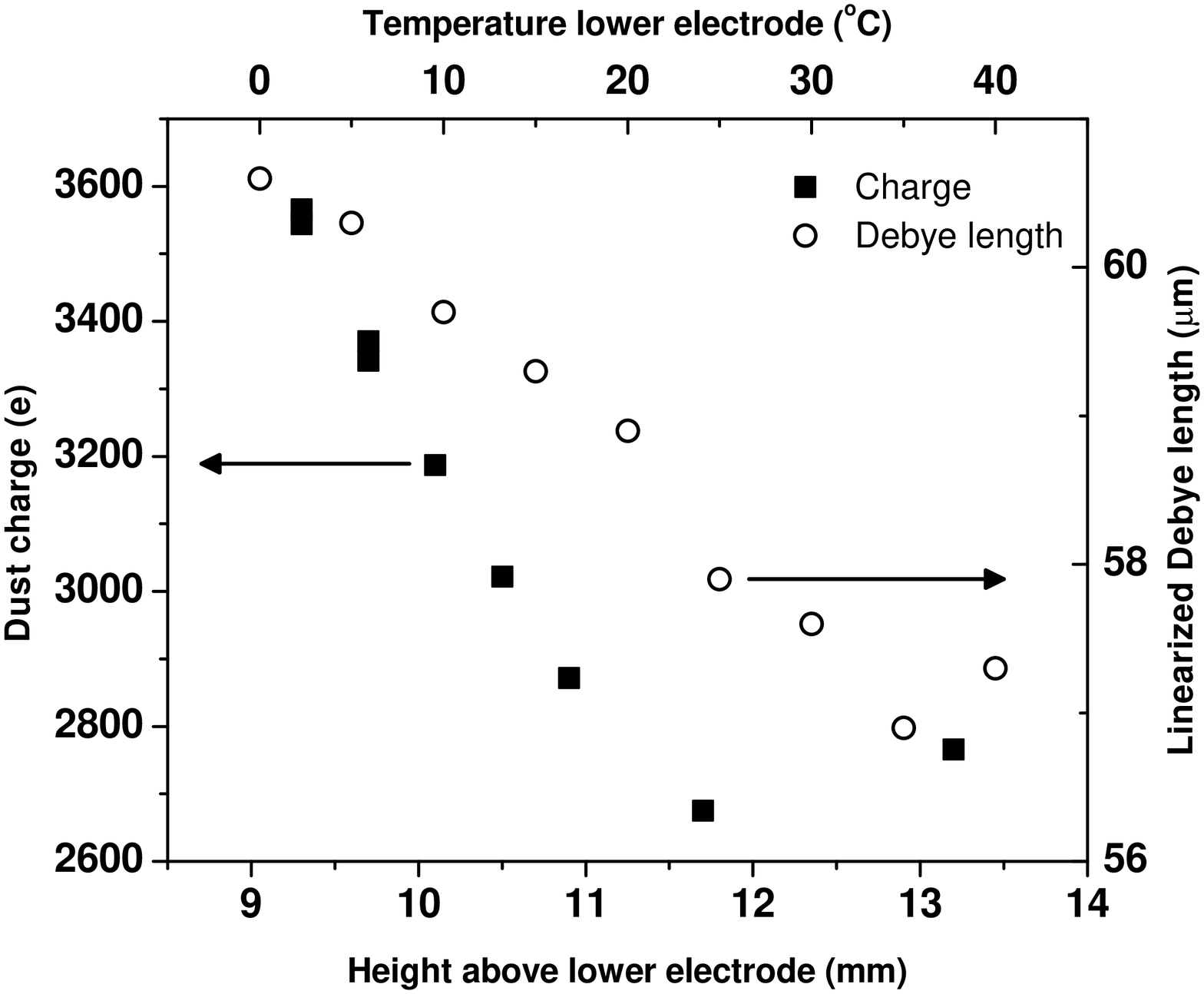}
\caption{The charge and Debye length above the lower electrode calculated with the fluid model discussed in the text. The dust charge is roughly 3000
electron charges and the Debye length is roughly 60 microns.}
\label{fig:fluid_qdld}
\end{figure}

Figure \ref{fit_electric_field} shows a fit of the electric field, obtained self-consistently with the fluid model. The
electric field in the sheath decreases linearly with height, whereas in the \emph{bulk}, the field can be well approximated
by a third order polynomial. For the mentioned experimental settings and geometry there is no extended quasi-neutral bulk 
where the electric field vanishes. The charge and Debye length obtained with the fluid model are shown in figure
\ref{fig:fluid_qdld}. Both linearly decrease with height, up to roughly 12 to 13 mm above the lower electrode, where an
increase in both is visible.

\begin{figure}[!ht]
\centering
\includegraphics[width=2.5in]{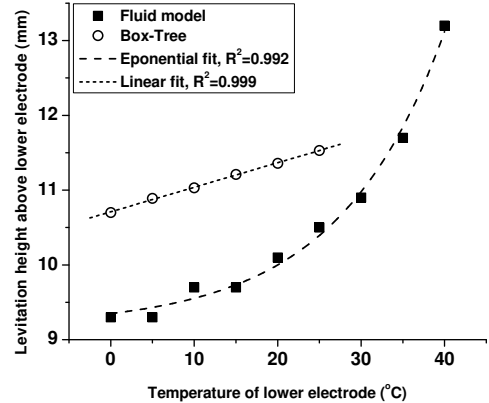}
\caption{The dust equilibrium levitation height obtained with the fluid model (solid squares) and box-tree (open circles).
The height varies exponentially with temperature according to the fluid model, but linearly for box-tree. This is to be expected, since
the ion drag is not included in the box-tree calculations.}
\label{levitation_heights}
\end{figure}

The equilibrium position of the dust fluid was also calculated while varying the electrode temperature. Figure
\ref{levitation_heights} shows the obtained levitation height above the electrode. Clearly, the levitation height
increases exponentially with applied temperature. This is consistent with the change in levitation height observed in
experiments, as is presented elsewhere in this issue \cite{part_2}.

\begin{figure}[!ht]
\centering
\includegraphics[width=2.5in]{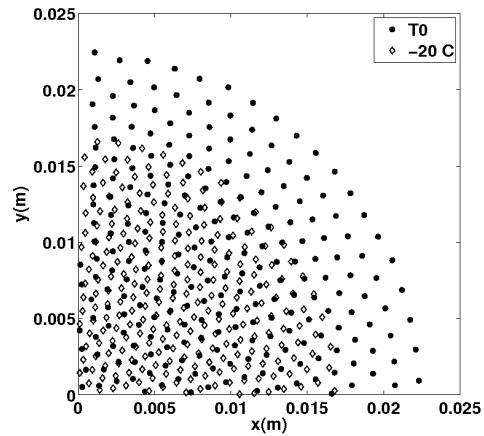}
\caption{Top-view of one quadrant of the dust crystal obtained in \emph{box-tree} atroom-temperature (black dots) and
with the lower electrode cooled to 20 degrees below room-temperature (diamonds).}
\label{crystals}
\end{figure}

The above results from the fluid model were used to prescribe the external forces in \emph{box-tree}, which was then run at each
temperature setting, until one minute real-time was simulated. At this
point the dust crystal had become stable. A top-view image of one quadrant of the dust crystals obtained for electrode
temperatures equal to room-temperature and 20${}^{\circ}$ below room-temperature are shown in figure \ref{crystals}.
Clearly, the crystal expands with increasing temperature, as would be expected from the behavior of $k$ in equation \ref{eq:keq}.

The dust levitation height obtained with \emph{box-tree} is also shown in figure \ref{levitation_heights}. The height
increases linearly with the temperature, rather than exponentially, as for the fluid model results. The primary difference
between the models is the missing ion drag calculation in \emph{box-tree}, impying that the ion drag \emph{is} important for the levitation of 2 micron particles in the plasma.

\begin{figure}[!ht]
\centering
\includegraphics[width=2.5in]{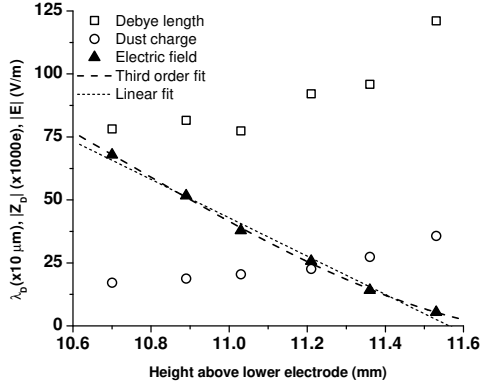}
\caption{The Debye length, dust charge, and vertical electric field obtained with \emph{box-tree}, together with a linear fit
and a third-order polynomial fit to the electric field.}
\label{final_results}
\end{figure}

The reconstructed Debye length, dust charge and electric field are shown in figure \ref{final_results} as a function of the
height above the lower electrode. The Debye length
increases from 750 $\mu$m to almost 1.3 mm, while the negative dust charge ranges from -17,000 to -34,000
electron charges. The electric field changes from -75 V/m to -7 V/m and can be better fitted by a third-order polynomial than
a straight line, indicating that the dust is not levitated in the region where the field varies linearly.

\section{Discussion and Conclusion}
By combining a self-consistent fluid model and an N-body code, the charge, Debye length and electric field in the sheath can be
reconstructed on sub-mm lengthscales. The ion drag clearly plays an important role, as evident from the linear increase
in levitation height in \emph{box-tree} where the ion drag is neglected, versus exponential increase in the fluid model, which self-consistently calculates
the ion drag force.

The values given for the Debye length and the dust charge are overestimated probably from an incorrect value of the radial confinement, 
through the calculation of $k$, since this confinement depends on the ion drag force. Compared to the fluid model, the reconstructed Debye
length is too large by a factor of 7, whereas the dust charge is overestimated by a factor of 5. This leads to an
underestimated electric field. Still, even if we multipily the electric field by this factor, the ion drift velocity
$u_{+} = {\mu}_{+}E$ is much smaller than the Bohm velocity, $u_B = \sqrt{kT_e/m_{+}}$. This implies that the dust
particles are levitated on the plasma side of the Bohm point, and in that sense above the sheath. This is also clear from
the fit to the electric field; the dust is levitated at a height where the electric field can already be approximated by a
third order polynomial, rather than in the region of linear electric field, which we would identify as the sheath.

Overall, the method shows promise. Adding the ion drag force in
\emph{box-tree} is a necessary extension, both for the vertical equilibrium height, as well as to obtain the proper radial
confinement. Using thermophoresis in a dusty discharge with two distinct particle sizes also provides a promising prospect:
since the thermophoretic force depends on the particle size, the vertical shift will be different for two crystals of
different size. By applying thermophoresis, it thus might be possible to squeeze two crystals together, study the interaction between
the dust clouds and any possible change in crystal properties. Since this could be achieved without changing the plasma properties directly, 
it would remove some of the ambiguity involved in analyzing dusty plasma systems.

\section*{Acknowledgment}

This research was made possible by NSF grant PHY-0648869 and NSF CAREER grant PHY-0847127, and we appreciated the useful discussions with
Dr. K. Qiao.

\vspace{-1.1cm}
\begin{IEEEbiography}[{\includegraphics[width=1in,height=1.25in,clip,keepaspectratio]{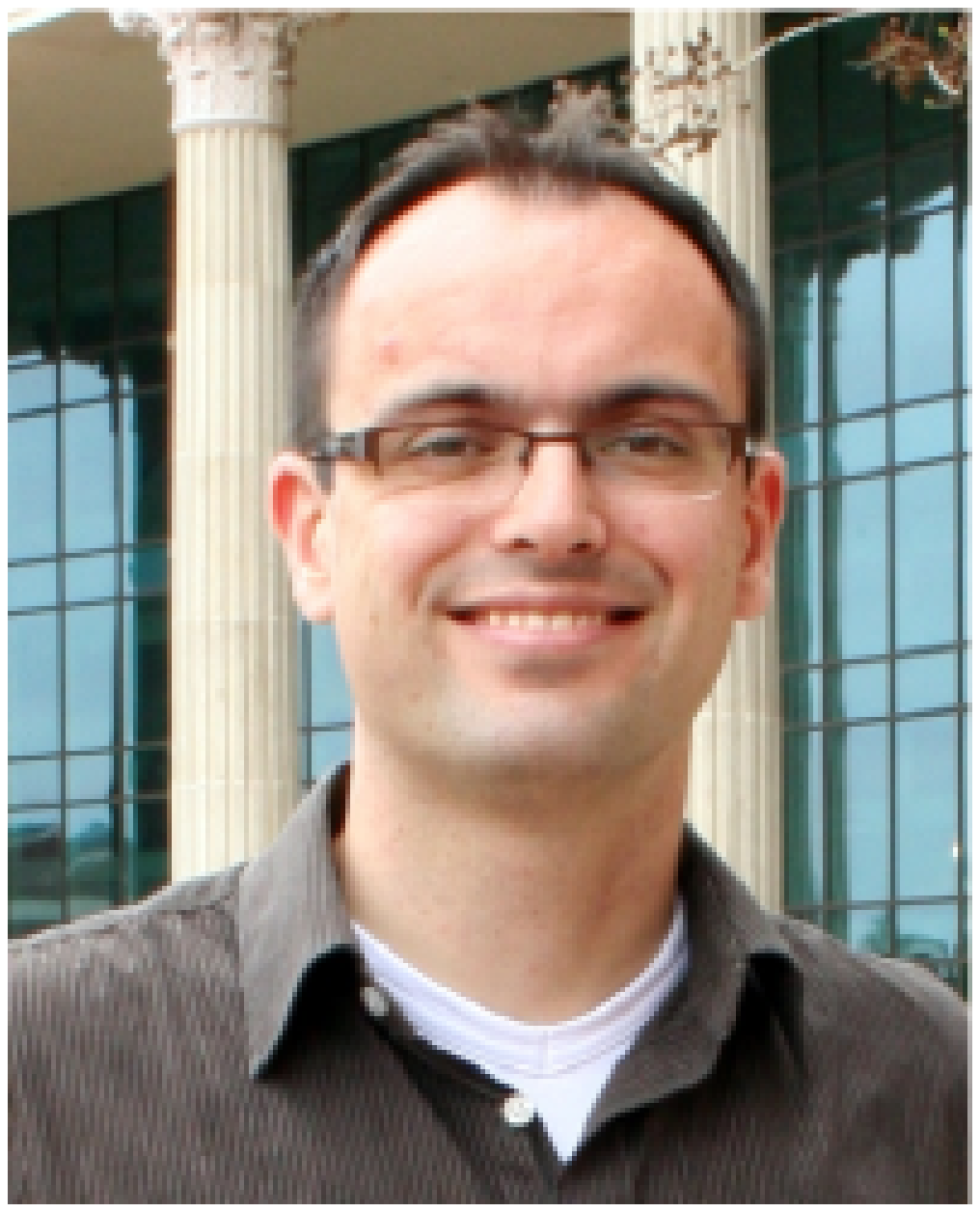}}]{Victor Land} was born in Petten, the Netherlands in 1979 and received his MSc in general astrophysics at Utrecht University in
the Netherlands in 2003, and his PhD at the
FOM-Institute for Plasma Physics 'Rijnhuizen', in the Netherlands in 2007. He is currently a post-doctorate research
associate at the Center for Astrophysics, Space Physics and Engineering Research at Baylor University, in Waco, Texas, where
he works on particle transport, charging, and coagulation in dusty plasma. 
\end{IEEEbiography}

\vspace{-1.4cm}
\begin{IEEEbiography}[{\includegraphics[width=1in,height=1.25in,clip,keepaspectratio]{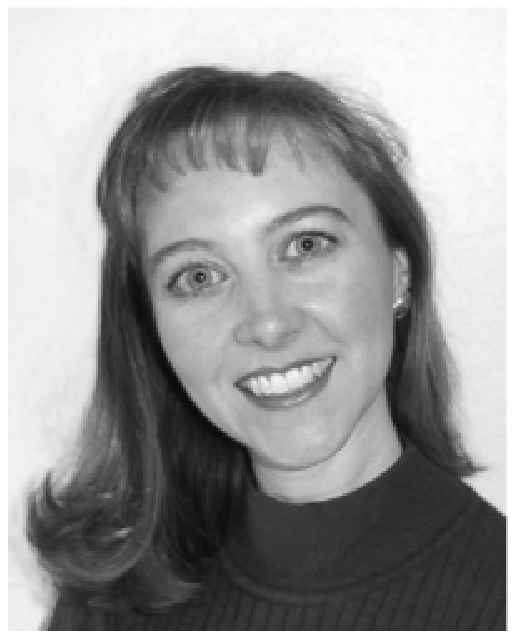}}]{Lorin Matthews}
was born in Paris, TX in 1972. She
received the B.S. and the Ph.D. degrees in physics from
Baylor University in Waco, TX, in 1994 and 1998,
respectively.
She is currently an Assistant Professor in the Physics
Department at Baylor University. Previously, she worked
at Raytheon Aircraft Integration Systems where she was
the Lead Vibroacoustics Engineer on NASA's SOFIA
(Stratospheric Observatory for Infrared Astronomy) project.
\end{IEEEbiography}

\vspace{-1.2cm}
\begin{IEEEbiography}[{\includegraphics[width=1in,height=1.25in,clip,keepaspectratio]{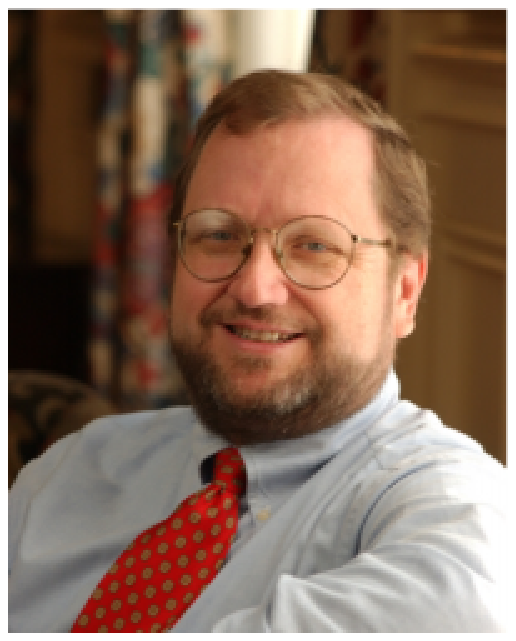}}]{Truell Hyde}
was born in Lubbock, Texas in 1956. He
received the B.S. in physics and mathematics from
Southern Nazarene University in l978 and the Ph.D. in
theoretical physics from Baylor University in 1988.
He is currently at Baylor University where he is the
Director of the Center for Astrophysics, Space Physics \&
Engineering Research (CASPER), a Professor of physics
and the Vice Provost for Research for the University. His
research interests include space physics, shock physics and waves and
nonlinear phenomena in complex (dusty) plasmas.
\end{IEEEbiography}

\end{document}